\documentclass[reprint,amsmath,amssymb,aps,]{revtex4-1}
\usepackage[mathlines]{lineno}
\usepackage{hyperref}
\usepackage{amsmath,graphicx}
\DeclareMathOperator{\evsym}{}
\newcommand\ev[1]{\evsym\left\langle#1\right\rangle}

\begin{document}

\preprint{APS/123-QED}

\title{Piezoelectrically Tuned Multimode Cavity Search for Axion Dark Matter}


\author{C. Boutan}
	\email[Correspondence to:]{christian.boutan@pnnl.gov}
  \affiliation{Pacific Northwest National Laboratory, Richland, WA 99354, USA}
  \author{M. Jones}
  \affiliation{Pacific Northwest National Laboratory, Richland, WA 99354, USA}
    \author{B. H. LaRoque}
  \affiliation{Pacific Northwest National Laboratory, Richland, WA 99354, USA}
  \author{N. S. Oblath}
  \affiliation{Pacific Northwest National Laboratory, Richland, WA 99354, USA}
  
\author{R. Cervantes}
  \affiliation{University of Washington, Seattle, WA 98195, USA}
\author{N. Du}%
  \affiliation{University of Washington, Seattle, WA 98195, USA}
  \author{N. Force}
  \affiliation{University of Washington, Seattle, WA 98195, USA}
    \author{S. Kimes}
  \affiliation{University of Washington, Seattle, WA 98195, USA}

      \author{R. Ottens}
  \affiliation{University of Washington, Seattle, WA 98195, USA}
 \author{L. J Rosenberg}%
  \affiliation{University of Washington, Seattle, WA 98195, USA}
  \author{G. Rybka}%
  \affiliation{University of Washington, Seattle, WA 98195, USA}
    \author{J. Yang}%
  \affiliation{University of Washington, Seattle, WA 98195, USA}

\author{G. Carosi}
\affiliation{Lawrence Livermore National Laboratory, Livermore, CA 94550, USA}
\author{N. Woollett}
\email[Correspondence to:]{woollett2@llnl.gov}
\affiliation{Lawrence Livermore National Laboratory, Livermore, CA 94550, USA}

\author{D. Bowring}
  \affiliation{Fermi National Accelerator Laboratory, Batavia IL 60510, USA}
\author{A. S. Chou} 
  \affiliation{Fermi National Accelerator Laboratory, Batavia IL 60510, USA}
    \author{R. Khatiwada}
  \affiliation{Fermi National Accelerator Laboratory, Batavia IL 60510, USA}
\author{A. Sonnenschein} 
  \affiliation{Fermi National Accelerator Laboratory, Batavia IL 60510, USA}
  \author{W. Wester} 
  \affiliation{Fermi National Accelerator Laboratory, Batavia IL 60510, USA}

\author{R. Bradley}
\affiliation{National Radio Astronomy Observatory, Charlottesville, VA 22903, USA}

\author{E. J. Daw}
\affiliation{University of Sheffield, Sheffield S3 7RH, United Kingdom}

\author{A. Agrawal}
\affiliation{University of Chicago, IL 60637}
\author{A. V. Dixit}
\affiliation{University of Chicago, IL 60637}

\author{J. Clarke}
  \affiliation{University of California, Berkeley, CA 94720, USA}
\author{S. R. O'Kelley}
  \affiliation{University of California, Berkeley, CA 94720, USA}
 
 \author{N. Crisosto}
  \affiliation{University of Florida, Gainesville, FL 32611, USA}
\author{J.~R.~Gleason}
  \affiliation{University of Florida, Gainesville, FL 32611, USA}
\author{S. Jois}
  \affiliation{University of Florida, Gainesville, FL 32611, USA}
 \author{P. Sikivie}
  \affiliation{University of Florida, Gainesville, FL 32611, USA}
\author{I. Stern}
  \affiliation{University of Florida, Gainesville, FL 32611, USA}
\author{N. S. Sullivan}
  \affiliation{University of Florida, Gainesville, FL 32611, USA}
\author{D. B. Tanner}
  \affiliation{University of Florida, Gainesville, FL 32611, USA}
  
   \author{P. M. Harrington}
  \affiliation{Washington University, St. Louis, MO 63130, USA}
    
  \author{E. Lentz}
  \affiliation{University of G\"{o}ttingen, G\"{o}ttingen 37077, Germany}

  \collaboration{ADMX Collaboration}\noaffiliation

\date{\today}

\begin{abstract}
The $\mu$eV axion is a well-motivated extension to the standard model. The Axion Dark Matter eXperiment (ADMX) collaboration seeks to discover this particle by looking for the resonant conversion of dark-matter axions to microwave photons in a strong magnetic field. In this paper we report results from an pathfinder experiment, the ADMX ``Sidecar'', which is designed to pave the way for future, higher mass, searches. This testbed experiment lives inside of and operates in tandem with the main ADMX experiment. The Sidecar experiment excludes masses in three widely spaced frequency ranges (4202-4249\,MHz, 5086-5799\,MHz and 7173-7203\,MHz). In addition, Sidecar demonstrates the successful use of a piezoelectric actuator for cavity tuning. Finally, this publication is the first to report data measured using both the TM$_{010}$ and TM$_{020}$ modes.
\end{abstract}

\pacs{Valid PACS appear here}
\maketitle




 Axions must exist in nature if the Strong CP problem, a vexing mystery within the Standard Model of particle physics, is solved by the existence of a spontaneously broken Peccei-Quinn symmetry \cite{Peccei:1977hh,Weinberg:1977ma,Wilczek:1977pj}. The fact that axions are non-baryonic, and can be made in sufficient abundance during the big bang, makes them attractive candidates for cold dark matter, an elusive, exotic, and weakly interacting form of matter that is approximately 5 times more prevalent than the baryonic matter in our universe \cite{Planck}. At small masses, the axion could account for a large fraction of the dark matter density \cite{Abbott:1982af,Preskill:1982cy,PhysRevLett.50.925}; cosmological considerations indicate that a likely mass range for which axions are the bulk of dark matter is $\mu$eV $< m_a <$ meV \cite{Bonati2016,PhysRevD.92.034507,Borsanyi2016,PhysRevLett.118.071802,PhysRevD.96.095001,PETRECZKY2016498}. The axion has the same quantum numbers as a neutral pion and the standard QCD axion has a fairly well defined relationship between its mass and coupling. Relaxing the requirement to solve the Strong CP problem allows for the existence of axion-like-particles (ALPs) with potentially larger couplings for a given mass \cite{ALPS}.

	Given that the dark matter density within our own Milky Way halo is expected to be $\rho_{a} \approx $ 0.45\,GeV/cm$^{3}$ \cite{DMdensity}, dark matter axions would have a local number density $\sim 10^{14}/$cm$^{3}$, but would remain almost impossible to detect due to their weak coupling. To date, the most promising detection scheme is the ``axion haloscope'' which exploits the inverse Primakoff effect \cite{Sikivie:1983ip}. In this scheme, a high-Q, tunable microwave cavity is immersed in a strong magnetic field. Dark matter axions interact with the static magnetic field, convert to photons, and deposit energy into the resonant mode of the cavity with a power

\begin{multline}
P_{a} = 1.14 \times 10^{-26}\,\text{W}  \left(\frac{g_{\gamma}}{0.97}\right)^{\!2}  \!\frac{\rho_{a}}{0.45\,\text{GeV}/\text{cm}^{3}} \left(\!\frac{B}{1 \, \text{T}}\!\right)^{\!2} \cdot  \\
\frac{V}{1 \, \ell }\cdot \frac{f}{1\,\text{GHz}} \cdot \frac{Q_{L}}{10,000} \cdot \frac{C_{mnp}}{0.5} ,
\label{eqn:pa}
\end{multline}

\noindent
where $g_{\gamma}$ is a model dependent constant ($-$0.97 for KSVZ \cite{Kim:1979if,Shifman:1979if} and +0.36 for DFSZ \cite{Dine:1981rt,Zhitnitsky:1980tq} axions), $B$ is the vertically oriented magnetic field strength, $V$ is the volume of the cavity, $f$ is the resonant frequency of the mode, $Q_{L}$ is the loaded quality factor of the resonant mode, and $C_{mnp}$ is an axion-photon conversion efficiency, commonly referred to as the form factor. The form factor, a number between 0 and 1, is defined as the overlap between the microwave electric field and the static magnetic field

\begin{align}
  C_{mnp}\equiv\frac{\lbrack \int_{V}dV\,\mathbf{E}_{mnp}(\mathbf{x},t)\cdot\mathbf{B(\mathbf{x})}\rbrack ^2}{ V B^2\int_{V}dV\, \epsilon _{r}{E_{mnp}}^{2}} ,
\label{eq:ff}
\end{align}

\noindent where $\mathbf{E}(\mathbf{x},t)$ is the electric field of the cavity mode, $\mathbf{B(x)}$ is the externally applied magnetic field, and $\epsilon _{r}$ is the relative permittivity within the cavity.  Given an empty cavity geometry and direction of $\mathbf{B(x)}$, only the electric field distribution of TM$_{0n0}$ modes result in non-zero form factors given by

\begin{align}
  C_{0n0}\equiv4/\chi^{2}_{0n} ,
\label{eq:ff}
\end{align}

\noindent where $\chi_{0n}$ is the n$^{th}$ zero of the Bessel function $J_{0}(x)$ \cite{Sikivie1985}. Because the form factor decreases for increasing $n$, previous haloscope searches have concentrated only on the TM$_{010}$ mode. However, analytically-calculated form factors only apply to the empty-cavity geometry. 3D electromagnetic field simulations which capture the complex geometry of a cavity containing a tuning rod have been shown to have an appreciable TM$_{020}$ form factor over a part of the tuning range. The peak magnitude of the TM$_{020}$ is significantly less than the TM$_{010}$ mode, which remains the discovery channel for the search.

When the frequency of the mode matches that of the axion, the power from the axion conversion will be given by Eq.\ref{eqn:pa}. The mode frequency, $f$, must be tuned to explore the possible range of axion masses because an axion haloscope is sensitive to axions within a narrow bandwidth, $f/Q$.


Over the past two decades, the Axion Dark Matter eXperiment (ADMX) has implemented the haloscope method described above, and has recently excluded the mass range 2.66--2.81\,$\mu$eV with DFSZ sensitivity \cite{PhysRevLett.120.151301}. Ultimately ADMX will scan up to axion masses of 40\,$\mu$eV. In this Letter, we present the results of a prototype higher-frequency axion search, the ADMX ``Sidecar'' \cite{thesisBoutan}. This pathfinder experiment was designed to pave the way for future, higher-mass ADMX searches without the need for an additional magnet or cryogenic system. In tandem with the primary lower frequency ADMX search \cite{PhysRevLett.120.151301}, the Sidecar cavity was operated in the 4--7\,GHz frequency range where it was sensitive to previously unexplored couplings of axion-like dark matter particles. Additionally, we demonstrate the readout from multiple modes and the use of piezoelectric tuning with the Sidecar search, paving the way for an ADMX search in this frequency range with DFSZ sensitivity.

\begin{figure}[htb!]
	\centering
                \includegraphics[width=0.45\textwidth]{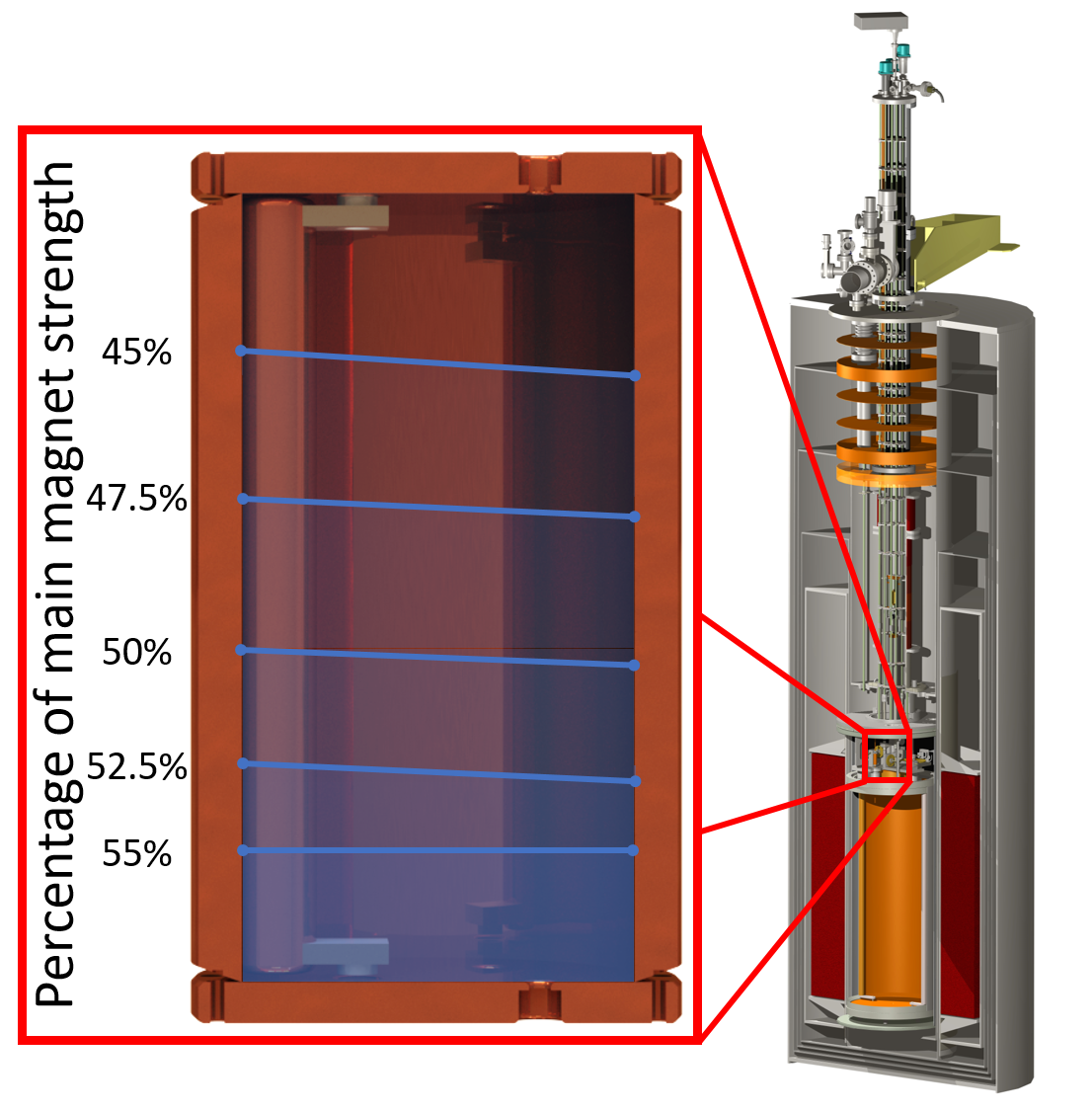}
                \caption{Schematic of the Sidecar cavity location in the ADMX insert. A simulation of the vertical component of the magnetic field in the volume of the cavity is shown overlaid. Contour lines indicate the field strength as a percentage of the center bore field strength.}
                \label{fig:schematic}
              \end{figure}

The Sidecar experiment resides inside the ADMX insert, an experimental apparatus that is lowered into a 3.4\,m cryostat housing a 8.5\,T superconducting magnet. The Sidecar cavity is placed directly above the ADMX main cavity within the magnet bore (Fig. \ref{fig:schematic}) where it is cooled to 150\,mK and experiences a field of 4.25\,T when the magnet is at full current. The 0.38\,$\ell$ OFHC copper plated stainless steel Sidecar cavity was fabricated at Lawrence Livermore National Laboratory. The cavity is formed with two end caps bolted to a 0.121\,m tall, 0.064\,m inner diameter cylindrical barrel with knife-edged ends to lower ohmic losses which would spoil the $Q$. The resonator is tuned with a 0.013\,m diameter copper-plated tuning rod, which is held parallel to the cylinder axis with a rotatable armature. The armature holds the rod 0.013\,m from the axis of rotation, allowing the rod to be revolved from the wall of the cavity to the center. The cavity and rod dimensions were designed to achieve a TM$_{010}$ frequency in the range 4--6\,GHz. At room temperature, the Sidecar cavity has an unloaded $Q$ of 22,000 when empty, and 11,000 with the tuning rod present. 

\begin{figure}[htb!]
	\centering
                \includegraphics[width=0.5\textwidth]{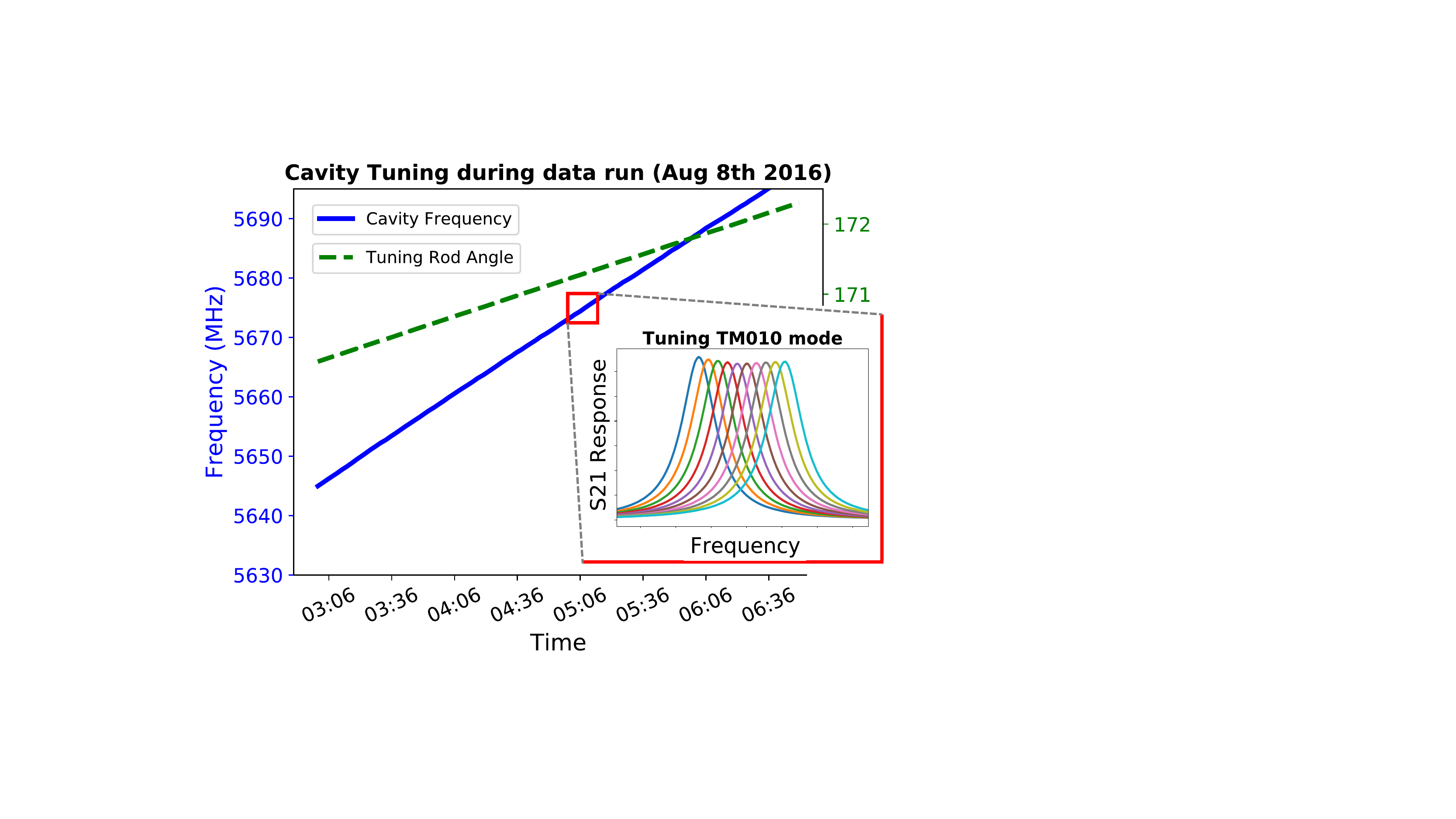}
                \caption{The cavity resonance shifted by positioning the tuning rod with an attocube piezoelectric actuator. The data shown was taken at 400\,mK during the data run B on August 8th, 2016. Roughly 1000 piezo motor steps were taken, resulting in a $\sim$2$^{\circ}$ revolution of the rod and 50\,MHz shift in the TM$_{010}$ mode.}
                \label{fig:piezo}
              \end{figure}

Achieving mechanical motion in a cryogenic space is a challenging aspect of this type of experiment. Unlike the main ADMX search, which communicates room temperature stepper motor motion to cryogenic gearboxes via long G-10 fiberglass shafts, the Sidecar system utilizes piezoelectric motors to directly drive the tuning rods. The stick-slip piezoelectric motors are made by attocube \cite{attocube2} and specified to operate at 10\,mK, 31\,T, and high vacuum simultaneously. The attocube ANR240/RES rotary and ANPz101eXT12/RES linear motors are mounted to the top end cap of the cavity and are used for tuning rod control and antenna depth adjustment. Figure\,\ref{fig:piezo} shows the effect of the rotary piezo actuation of the tuning rod on the Sidecar TM$_{010}$ mode as seen in a series of cavity swept response measurements. The cavity frequency is shown to smoothly tune 50\,MHz by slowly revolving the rod $\sim$2$^{\circ}$, over the course of several hours.

The cavity antenna was made by stripping a section of SMA-terminated, semi-rigid coax. Its coupling to the cavity is measured by directing a swept signal through a circulator towards the antenna. In the case where the antenna is critically coupled to the resonant mode, all power is absorbed into the cavity on resonance. Under normal conditions, the linear piezo motor, with a maximum travel length of 12\,mm adjusts the depth of the antenna to maintain a return loss of less than -20\,dB relative to the off-resonance reflected baseline.  

When tuning the cavity resonance, minimization of the piezoelectric heat load is necessary to maintain milli-kelvin temperatures. The piezo motors have three sources of heating: dissipated power from exercising the actuator, mechanical heating from the stick-slip motion, and ohmic heating from the finite resistance of the piezo element. The latter two terms, heating from friction and leakage currents, were found to be subdominant compared with the power dissipated by the actuator \cite{AttocubeHeating}. The dominant source of heating is expressed as

\begin{equation}
	\begin{split}
		P_{\text{dissipated}} &= 2CV^{2}_{p}\tan(\delta)f_{d} \\
        &\sim 2(200 \, \textrm{nF})(60\, \textrm{V})^{2}(0.02)f_{d} \\
		&\sim (30\,\mu \textrm{J}\text{ per step}) \times (\text{steps per second}),
	\end{split}
\end{equation}

\noindent where $C$ is the capacitance of the actuator, $V_{p}$ is the peak voltage applied, tan($\delta$) is the loss tangent, and $f_{d}$ is the driving frequency of the sawtooth voltage applied to the motor. Using typical parameter values, the theoretical heating of the rotary motor is estimated to be $\sim$\,50\,mJ/degree, and that of the linear antenna motor to be $\sim$\,100\,mJ/mm. However, under normal operating conditions, the linear motor is stationary and the rotary motor takes approximately 5 steps per minute, resulting in an average heating of less than 5\,$\mu$W. Although the cavity and piezo motors are thermally tied to the mixing chamber of a dilution refrigerator, with 800$\mu$W cooling capacity, motor heating causes a negligible increase in temperature. 

The cavity output is coupled to a Low Noise Factory HFET amplifier LNF-LNC1$\_$12A which is thermally sunk to the top of the helium reservoir. These amplifiers have a characteristic stable broadband gain of 40\,dB, and a characteristic noise temperature at or below 4\,K. \textit{In-situ} noise calibration tests were achieved with a cryogenic microwave switch that toggled the input to the amplifier from the cavity to a heatable 50 $\Omega$ termination. 
	
The room-temperature portion of the receiver is a single heterodyne design, mixing the 4--7\,GHz cavity frequency down to 10.7\,MHz. The signal is also further amplified and filtered before it is digitized and undergoes a secondary stage of mixing in software. A cryogenic circulator placed directly after the cavity antenna, a Mini-Circuits RC-8SPDT-A18 radio frequency (RF) switch matrix, and a network analyzer allows the signal path to be reconfigured between data collection and measurements of cavity frequency, antenna coupling, and noise temperature.


The data-taking cadence consists of: (i) rotate the tuning rod to shift the relevant TM$_{0n0}$ mode by a fraction of its bandwidth, (ii) obtain the center frequency, $Q$ and antenna coupling from network analyzer measurements, (iii) set the receiver local oscillator frequency such that the resonant frequency is mixed down to 10.7\,MHz, and (iv) digitize for 100 seconds. This process is fully automated and runs in parallel with the main experiment. For each iteration, we measure the parameters needed to calculate the expected power in Eq.\,\ref{eqn:pa}. The frequency step size for each tuning motion is adjusted to obtain the integration time needed to reach a target sensitivity and signal-to-noise ratio. During data collection, RF power is occasionally injected into the cavity to simulate an axion signal. Observation of these ``synthetic axions'' confirmed the integrity of the signal path. Frequency regions with statistically large powers are rescanned to increase confidence in a potential axion signal.  In practice all large power regions became statistically insignificant with the additional data.

\begin{figure}[htb!]
	\centering
                \includegraphics[width=0.45\textwidth]{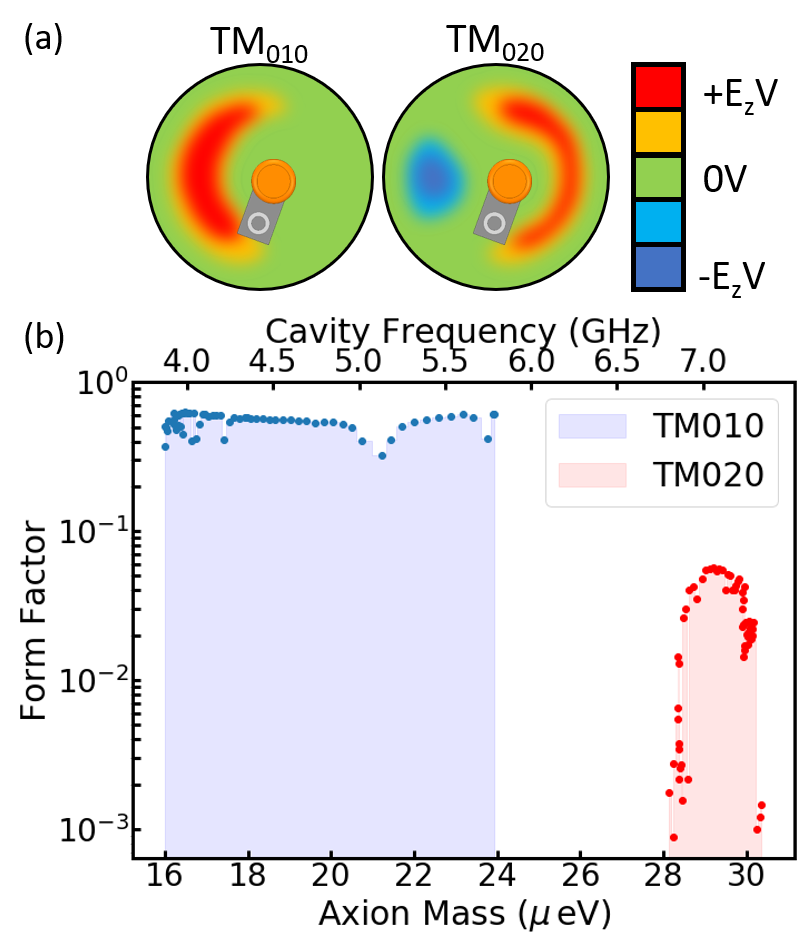}
                \caption{\label{fig:FormFactor} Electric field distributions and form factor values for different tuning rod positions. (a) The distribution of the electric fields of the TM$_{010}$ and TM$_{020}$-like modes are shown for a tuning angle of 160$^{\circ}$. (b) Calculated form factor of the TM$_{010}$ and TM$_{020}$ modes as a function of mode frequency. Choppy features where the form factor significantly decreases are caused by mode crossings at which the mode of interest tunes into and mixes with an interfering mode.}                
\end{figure}

The data run plan requires a careful study of the cavity resonant structure to calculate the form factor (Eq.\,\ref{eq:ff}). A map of the mode structure for a set of tuning rod locations ranging from the center to the wall was created by using an eigenmode solver to find the resonant frequencies over the tuning range of the TM$_{010}$ and TM$_{020}$ modes. Simulations performed in HFSS \cite{HFSS2} are used to identify the relevant mode and calculate its form factor as a function of rod position. Figure \,\ref{fig:FormFactor}b shows the form factor of the TM$_{010}$ and TM$_{020}$ like modes.

The axion search data used in this work were collected between August of 2016 and June of 2017 and are composed of three separate data sets labeled in order of increasing frequency and summarized in Table \ref{tab:RunSummary}. Run B is a low magnetic field commissioning run that emphasized testing the piezoelectric tuning capabilities of the experiment, rather than sensitivity. Several months later, a high frequency run (Run C) and a low frequency run (Run A) focused on achieving higher sensitivity over a narrower frequency range. The linear piezo motor responsible for antenna coupling failed early on, causing the loaded Q to be suboptimal and the antenna to remain in an over-coupled state for the duration of the run.

\begin{table}[h]
  \caption{Data Run Summary}
  \begin{tabular}{l*{6}{c}r}
  \label{tab:RunSummary}
  Run              	& A & B & C  \\
  \hline
  Timeline 		& May 24-June 11 &  Aug 9-Oct 4  & Feb 27-April 9   \\
     		& 2017 &  2016  & 2017   \\
  Mode 	& TM$_{010} $ & TM$_{010}$ & TM$_{020}$  \\
  Freq (MHz) 	& 4202 - 4249 & 5086 - 5799  & 7173 - 7203   \\
  Mass ($\mu$eV) 	& 17.38 - 17.57 & 21.03 - 23.98 & 29.67 - 29.79    \\
  Usable Spectra 	& 11k & 32k & 24k   \\
  \textit{B}-Field (T)		& 3.11 & 0.78 (2.55\footnote{The magnet was ramped to a higher field for 3 days just before the end of the data run}) & 3.11   \\
  Form Factor     	& 0.49  & 0.44--0.61  &  0.04--0.046  \\
  Q$_{L}$     	& 6.2k  & 2.2k  &  2.3k  \\
  T$_{sys}$ (K)     	& 7.0 $\pm$  0.3  & 7.0 $\pm$  0.3  &  7.1 $\pm$  0.3  \\
  \end{tabular}
\end{table}


\begin{figure*}[htb!]
  \centering
  \includegraphics[width=0.30\textwidth]{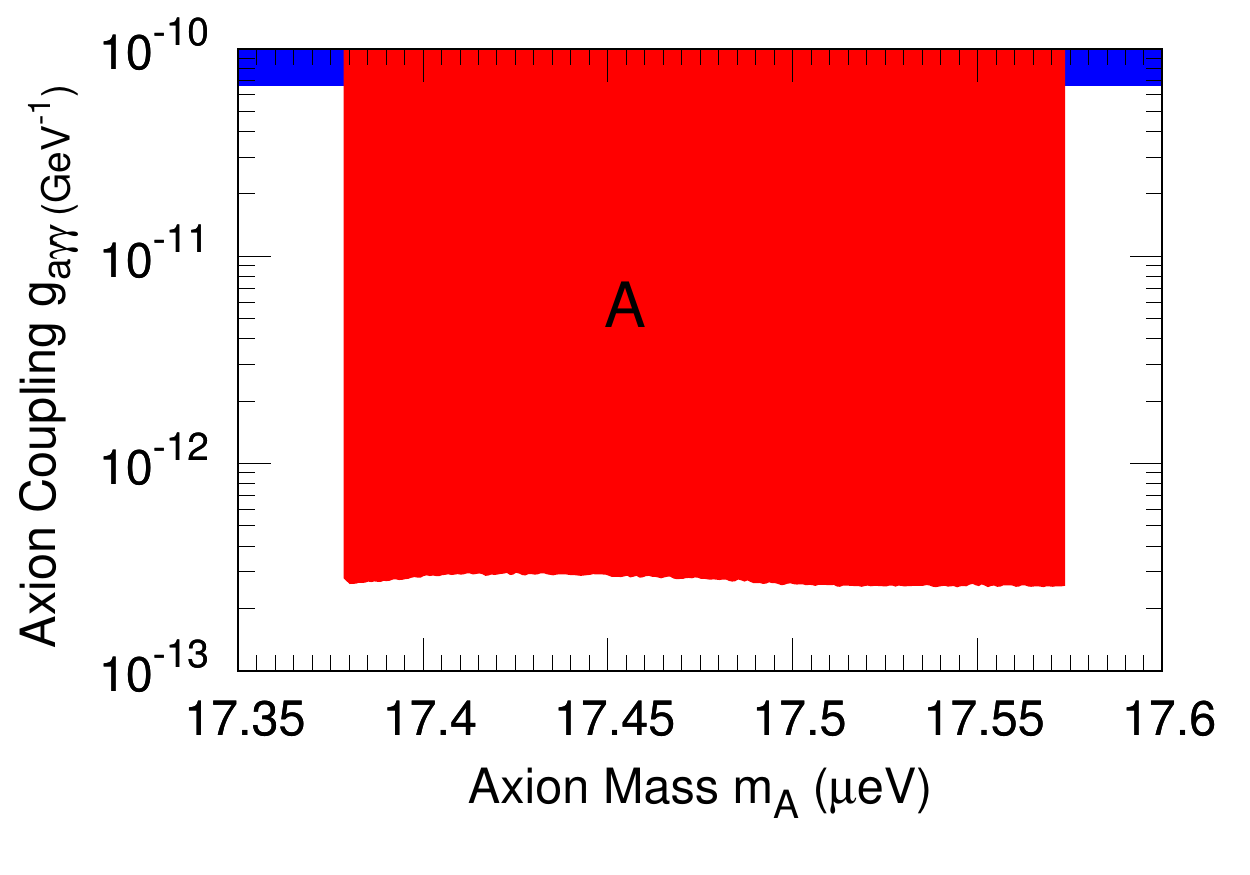}
  \includegraphics[width=0.30\textwidth]{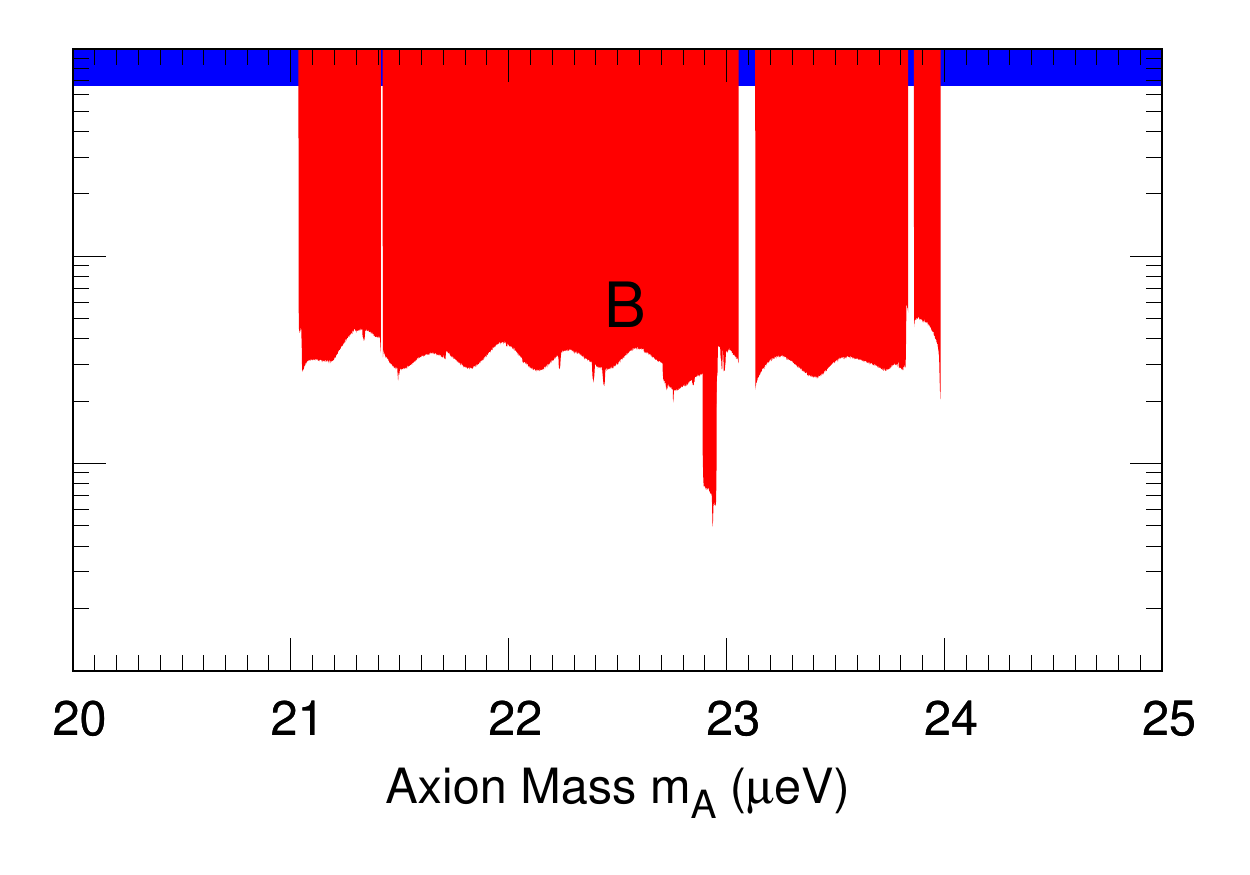}
  \includegraphics[width=0.30\textwidth]{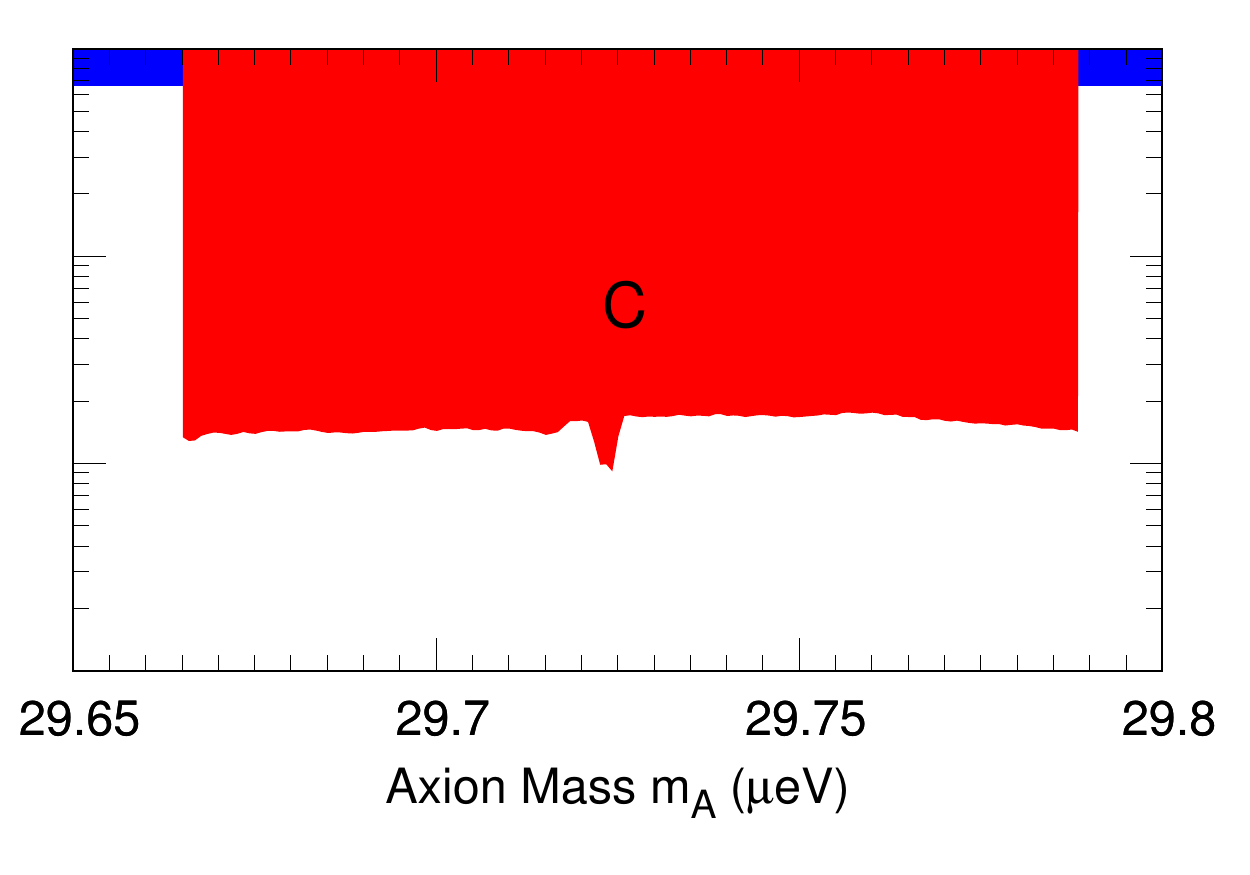}
  \includegraphics[width=0.91\textwidth]{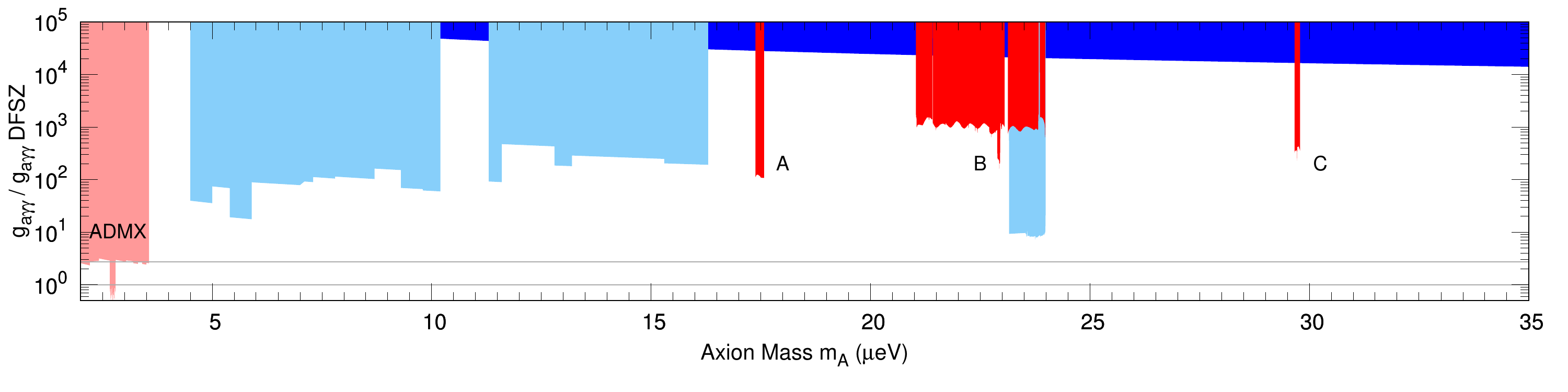}
  \caption{90$\%$ confidence exclusion on axion-photon coupling with the assumption of dark matter axions at a density of 0.45\,GeV/cm$^{3}$ with an isothermal distribution. \textit{Top:}  Zoomed-in limits set over the three frequency regions. Regions A and B were scanned using the TM$_{010}$ mode, while region C was scanned using the TM$_{020}$ mode.  \textit{Bottom:} Results from this work (red) shown in the context of the ADMX (light red)\cite{PhysRevD.69.011101,Asztalos:2009yp,PhysRevD.94.082001,Sloan201695,PhysRevLett.120.151301,PhysRevLett.80.2043,PhysRevD.64.092003,1538-4357-571-1-L27}, other axion haloscope searches (light blue) \cite{DePanfilis:1987dk,Wuensch:1989sa,Hagmann:1990tj,PhysRevLett.118.061302,PhysRevD.97.092001}, and the solar axion exclusion limit (dark blue) \cite{cast2017}. The DFSZ and KSVZ predictions for the coupling constant are the thin grey lines.}
  \label{fig:Limits}
\end{figure*}

The full process by which many raw spectra are combined and converted into limits on the axion-photon coupling ($g_{a\gamma \gamma}$) is described in \cite{Brubaker:2017rna}. The procedure begins by taking digitized spectra from the IF band of the receiver, centered at 10.7\,MHz, and correcting them for the receiver response. This is done by first dividing each spectrum by the time-independent response, achieved by averaging all of the raw spectra in the data set. Secondly, each spectrum is processed with a Savitsky-Golay filter (length 129 and polynomial order 2) to 95\% of the least-deviant power bins to characterize any slowly varying frequency specific receiver structure without fitting out a narrow $\sim$5\,kHz wide axion signal. Each residual spectrum is then divided by its fit and the baseline of 1 was subtracted to yield unitless spectra with zero mean. Axion-like signals were added to the data to measure the degree of sensitivity degradation caused by the fit.  A histogram of all of the residual spectra is well fit by a Gaussian function with $\mu$ = 0.01 and $\sigma$ = 0.98, in agreement with the expected probability distribution. The zero-mean, unitless spectra are then checked for external interference and finally scaled by the bin resolution, Boltzmann constant, and system temperature $T_{sys}$ to achieve a representation of real power fluctuations in the cavity. The system temperature is calculated by using the Friis amplifier noise temperature equation \cite{friis1944}

\begin{equation}
\label{eq:Friis}
T_{sys} = T_{phys} + T_{1} + \frac{T_{2}}{G_{1}} + \frac{T_{3}}{G_{1}G_{2}} +...,
\end{equation}
where $T_{phys}$ is the physical temperature, $T_{1}$ and $G_{1}$ are the noise temperature and gain of the first amplifier and so on. Due to a failure with one of the cryogenic switches, the noise temperature of the Low Noise Factory amplifier was not measured concurrently with the axion search data. It was measured \textit{in situ} with a functioning switch in a later run of the sidecar experiment, operating at a physical temperature of 4\,K. The system noise contribution from the receiver, including the first-stage amplifier, was calibrated by measuring the change in output rf power as the temperature of the heated load was changed, and found to be 4.0$\pm$0.3 K. This was consistent with, but slightly higher than, the manufacturer specification of a 3\,K amplifier noise at a physical temperature of 10\,K, and so we use our more conservative \textit{in situ} value. At the end of the single scan correction process, spectra represent thermal noise fluctuations that may contain a very small excess of power from an axion signal.

Corrected spectra are then weighted according to their sensitivity, so that data with different system noise, magnetic fields, loaded quality factors, and form factors could be directly summed. Thousands of overlapping and non-overlapping spectra are added in quadrature to form a single ``grand power spectrum''. For a typical isothermal velocity dispersion $\ev{v^{2}}^{1/2}$ = 270\,km/s, the expected axion signal width is approximately $\Delta f \cong m_{a}\ev{v^{2}}/h$, where $h$ is Planck's constant. The grand power spectrum is examined for the signs of an axion signature line shape using a dynamic linewidth filter as predicted by an isothermal distribution for a fixed axion density $\rho_{a}$ = 0.45\,GeV/cm$^{3}$. 

The total signal degradation factor from receiver structure removal and axion signal frequency bin misalignment was found to be $\eta$ = 0.78 by injecting hundreds of synthetic software signals in a Monte Carlo simulation. A 95$\%$ confidence threshold on the amount of measured power was set and new limits on $g_{a \gamma \gamma}$ were obtained as shown in Fig\,\ref{fig:Limits}.


In summary, the ADMX Sidecar has produced new limits on the axion coupling factor in three frequency bands within the range 4.2\,GHz--7.2\,GHz.  This is the first time an axion haloscope has been operated over much of this region and improves on the solar bounds \cite{CAST} by two orders of magnitude under the assumption of 100$\%$ dark matter axion-like particles. While this sensitivity is not sufficient to reach the axion coupling predicted for QCD axions, it excludes a meaningful $g_{a \gamma \gamma}$ parameter space for generically motivated axions, and spearheads experimental efforts for DFSZ sensitivity in this axion mass range.

Higher frequency cavities are required to explore higher axion masses. With Sidecar, it has been shown that smaller high frequency cavities can successfully scan over multiple modes, providing a much wider scan range without requiring physical changes to the system. The TM$_{020}$ mode suffers from a weaker form factor, leading to a factor of $\sim$4 loss in expected axion power but the exclusions produced are still significant.

Future runs of Sidecar will concentrate on covering more of the plausible axion mass range and adopting the use of quantum limited amplifiers. By placing a near-quantum-limited traveling wave parametric amplifier (TWPA) \cite{TWPA2} before the HFET, the system noise temperature for future data runs could be reduced by a factor of 10, resulting in a 100x scan rate improvement. Since these pre-amplifiers are broadband, unlike the MSA and JPA amplifiers used in the AMDX experiment, future Sidecar runs may take data on both the TM$_{010}$ and TM$_{020}$ modes simultaneously with a single cryogenic signal path. The signal would be divided at the 300 K heterodyne receiver, bandpass filtered, mixed down separately, and digitized to achieve a dual channel axion search. Sidecar will continue to operate in tandem with the main ADMX search, allowing for the \textit{in situ} testing of new haloscope techniques.

\section{Acknowledgments}

This work was supported by the U.S. Department of Energy through Grants Nos. DE-SC0009723, DE-SC0010296, DE-SC0010280, DE-SC0010280, DE-FG02-97ER41029, DE-FG02-96ER40956, DE-AC52-07NA27344, and DE-C03-76SF00098. This manuscript has been authored by Fermi Research Alliance, LLC under Contract No. DE-AC02-07CH11359 with the U.S. Department of Energy, Office of Science, Office of High Energy Physics. Additional support was provided by the Heising-Simons Foundation and by the LDRD offices of the Lawrence Livermore and Pacific Northwest National Laboratories. LLNL Release Number: LLNL- JRNL-763299.

\bibliographystyle{apsrev4-1}
\bibliography{2017admx_main}

\end{document}